\documentclass[sigconf]{acmart}

\usepackage[utf8]{inputenc}
\usepackage{bm}
\usepackage{balance}
\usepackage{booktabs} 
\usepackage{graphics}
\usepackage{multirow}
\usepackage{array}
\usepackage{flushend}
\usepackage[english]{babel}
\usepackage[T1]{fontenc}
\usepackage{textcomp}
\usepackage{latexsym}
\usepackage{natbib}
\usepackage{graphicx}
\usepackage{multirow}
\usepackage[labelformat=simple]{subcaption}
\usepackage{algorithm}
\usepackage{algorithmic}

\copyrightyear{2021}
\acmYear{2021}
\setcopyright{acmlicensed}\acmConference[SIGIR '21]{Proceedings of the 44th International ACM SIGIR Conference on Research and Development in Information Retrieval}{July 11--15, 2021}{Virtual Event, Canada}
\acmBooktitle{Proceedings of the 44th International ACM SIGIR Conference on Research and Development in Information Retrieval (SIGIR '21), July 11--15, 2021, Virtual Event, Canada}
\acmPrice{15.00}
\acmDOI{10.1145/3404835.3462918}
\acmISBN{978-1-4503-8037-9/21/07}
\begin{document}
\title{Group based Personalized Search by Integrating Search Behaviour and Friend Network}
\author{Yujia Zhou$^{2,3}$, Zhicheng Dou$^{1*}$, Bingzheng Wei$^{3}$, Ruobing Xie$^{3}$, and Ji-Rong Wen$^{4,5}$}

\affiliation{%
  $^1$Gaoling School of Artificial Intelligence, Renmin University of China \\
  $^2$School of Information, Renmin University of China \\
  $^3$WeChat Search Application Department, Tencent, China \\
  $^4$Beijing Key Laboratory of Big Data Management and Analysis Methods \\
  $^5$Key Laboratory of Data Engineering and Knowledge Engineering, MOE
 \country{}
}

\email{{zhouyujia, *dou}@ruc.edu.cn}

\begin{abstract}
The key to personalized search is to build the user profile based on historical behaviour. To deal with the users who lack historical data, group based personalized models were proposed to incorporate the profiles of similar users when re-ranking the results. However, similar users are mostly found based on simple lexical or topical similarity in search behaviours. In this paper, we propose a neural network enhanced method to highlight similar users in semantic space. Furthermore, we argue that the behaviour-based similar users are still insufficient to understand a new query when user's historical activities are limited. To tackle this issue, we introduce the friend network into personalized search to determine the closeness between users in another way. Since the friendship is often formed based on similar background or interest, there are plenty of personalized signals hidden in the friend network naturally. Specifically, we propose a friend network enhanced personalized search model, which groups the user into multiple friend circles based on search behaviours and friend relations respectively. These two types of friend circles are complementary to construct a more comprehensive group profile for refining the personalization. Experimental results show the significant improvement of our model over existing personalized search models.
\end{abstract}


\def\authors{Yujia Zhou, Zhicheng Dou, Bingzheng Wei, Ruobing Xie, Ji-Rong Wen}
\settopmatter{printacmref=true}
\maketitle

\section{Introduction}
\label{sec:intro}
Search engine is a common tool for obtaining information, but returning the same results to different users is often not optimal. Personalized search aims to customize ranking lists to meet individual needs for each user. It has been proven to be effective in improving user search experience \cite{cai2014personalized}. Previous studies \cite{bennett2012modeling, cai2014personalized, song2014adapting, teevan2011understanding, harvey2013building, Vu2014Improving, vu2015temporal} usually build user profiles based on historical behaviours, and consider the matching between the user profiles and documents when ranking the results. Recently, the emergence of deep learning allows models to capture user interests in semantic space \cite{hrnncikm, ZhouDW20, Lu:2019, li2014deep} and brings significant improvement in search quality.

Although the existing strategies of personalization are diversified, most of them point out that personalization is extremely dependent on user's historical behaviour \cite{dou2007large, teevan2011understanding}. When the user has only limited activities in history, the effect of personalized strategies is also constrained. To tackle this problem, some group based methods \cite{TeevanDH10, TeevanMB09, Vu2014Improving} were proposed to integrate query logs of similar users to expand user profiles. However, these studies mainly identify similar users based on lexical or topical similarity of queries and documents. This approach is too simple to inevitably introduce a lot of noisy users. In fact, even if users issue same queries, they may be driven by different intents. Besides, hot events will also encourage two users with different profiles to have similar search behaviours. In this paper, \textbf{we propose a better personalized method in highlighting similar users with the help of neural networks}. This enables us to calculate the similarity between users in semantic space and in a more reliable manner.

Previous approaches for finding similar users mainly refer to the user's historical search or click behaviours, and have demonstrated some effectiveness in group based personalized search. However, due to the reliance on the historical behaviour, these methods are still insufficient when the users only have limited historical data. It is unreliable to understand the intent of a new query based on similar users found only with scant query log. Recently, the integration of social platforms and search engines has gradually become a hot field. Many APPs, such as WeChat and Toutiao, have both social attributes and information acquisition functions. These APPs generate two types of connected data: search activities of individual users, and the friend relationship among users. This enables the possibility of leveraging friend network to improve personalized web search, and in this paper, \textbf{we carry out a preliminary study on introducing the user's friend network into personalized search} to solve the above-mentioned problem. Friend network has been applied on many user modeling tasks, especially on social recommendation \cite{WangLWF20, MuZHT19, Fan0LHZTY19}, and shows the effectiveness in personalization. In real life, the establishment of friendship is often based on the same background or interest. This will help capture the user’s stable profile based on his friends even if his query log is empty. Endowed with the benefit of friend network, the reliability of group formation is enhanced to address the problem of data sparsity.

Intuitively, not all friends contribute equally to building the user profile. Both the search behaviour similarity and friend relation closeness can determine their contribution to the current user in personalization. The former reflects the similarities between users' information needs, while the latter reveals the strength of relationship between users in the real world. Their complementary advantages can help build a more comprehensive user profile. In other words, if a close friend also has similar search behaviours as the current user, he should be given more attention in personalization. Based on this consideration, \textbf{we propose integrating both factors, behaviour similarity and friend relationship, in measuring the similarity between users}.

Furthermore, in real life, a user usually belongs to multiple friend circles due to his diverse backgrounds or interests. Users in the same circle generally share similarities in one aspect, such as belonging to the same university or loving to play football. This observation inspires us to group the user into multiple friend circles to build the group profile in a fine-grained way. The user's historical search behaviour and the friend network provide us with two grouping angles: one is based on similar information needs, and the other is based on the relationship in real life. To ensure the diversity of friend circles, we design an algorithm to maximize the difference of the users included in different circles. Thus, each friend circle reflects one aspect of the user profile, and they together form a complete one for better personalization.

Inspired by the above observation, we propose a group based personalized search model FNPS, which integrates the search behaviour and friend network to model the group profile with neural networks. The construction of group profile can be summarized as the following four steps. \textbf{Firstly}, we leverage the friend network and the user's historical search behaviours to form friend circles from two angles. \textbf{Secondly}, the graph attention network is applied to aggregate the friends' profiles with different weights. \textbf{Thirdly}, the representations of two types of friend circles are fed into a cross attention layer to enhance each other. \textbf{Finally}, in response to the current query intent, we use the query-aware attention to highlight relevant friend circles and construct the group profile dynamically. The final personalized results are generated under the combined effect of the user's individual profile and the group profile.

Our main contributions can be summarized as follows. (1) To the best of our knowledge, this is the first time that similar users are used to build profiles for neural personalized web search; (2) In addition to the user similarity calculated based on users' search and click behaviour, we introduce the friend network into personalized search with neural networks to further tackle the problem of historical data sparsity. (3) Empowered by the combination of search behaviour similarity and friend relation closeness, the measurement of similarity between users is more reliable. (4) To build the group profile in a fine-grained way, we group the user into different friend circles from two angles. They will be considered with dynamic weights based on the current query in personalization.

\section{Related work}
\label{sec:related}
\subsection{Personalized Web Search}
The goal of personalized search is to re-rank the results to match the individual needs for different users. The key to personalized search is how to build user profiles based on their search logs. Some early studies mainly construct user profiles by extracting features from query logs, such as click features and topic features. Dou et al. \cite{dou2007large} and Teevan et al. \cite{teevan2011understanding} predicted the click probability of a document by counting the number of historical clicks. They pointed out that it was reliable to personalize search results with respect to user's re-finding behaviours. The extraction of topic features has undergone a transition from manual design to automatic learning \cite{bennett2010classification, sieg2007web, white2013enhancing}. Due to the huge human cost of the former, some studies \cite{Carman2010Towards, harvey2013building, vu2015temporal, vu2017search} focused on how to automatically learn the topic representation of the document with Latent Dirichlet Allocation (LDA). With the emergence of the learning to rank method, many studies \cite{bennett2012modeling, volkovs2015context, white2013enhancing} combined multiple personalized features for ranking with LambdaMART \cite{wu2008ranking} algorithm.

Recently, deep learning has been applied in various research fields and has achieved success. Due to its powerful representation learning capabilities, it is good at extracting user potential interests in personalized search. Song et al. \cite{song2014adapting} designed a deep neural network to adapt a generic RankNet for personalized search. Ge et al. \cite{hrnncikm} focused on sequential information hidden in query log and learned the user interests with recurrent neural network. Zhou et al. \cite{ZhouDW20} built memory networks to generalize the re-finding behaviour to semantic space. Lu et al. \cite{sigir/LuDXWW20} introduced knowledge base to refine user profiles. Moreover, methods based on adversarial neural networks \cite{Lu:2019} and reinforcement learning \cite{YaoDXW20} have also been proposed for enhancing data quality. In addition to these user profile-based methods, Zhou et al. \cite{sigir/ZhouDW20} and Yao et al. \cite{sigir/YaoDW20} argued that the query representation is dynamically changing in different historical contexts. They used the history to learn the embedding of the current query. This paper attempts to introduce the friend network into personalized search with deep learning to enhance user profiles, especially for the users who lack historical data.

\subsection{Group based Web Search}
Group based search aims to improve search results with the help of similar users' query logs. The existing methods for extracting similar users can be divided into two categories, based on search behaviour or social relations. For the first method, Dou et al. \cite{dou2007large} proposed the G-Click model to find top K users with similar search behaviour, and then ranked the results based on these users. Morris et al. \cite{MorrisH07} focused on enhancing the quality of traditional web search with collaborative behaviours. Teevan et al. \cite{TeevanDH10, TeevanMB09} tried different grouping ways and showed that grouping could identify what users would consider relevant to the query. However, these static grouping methods ignore the users' diverse interests with respect to different topics. Vu et al. \cite{Vu2014Improving} proposed to construct the group dynamically in response to the input query.

The social network based methods were proposed to model user preferences based on social relations. Bender et al. \cite{BenderCKMNPSW08} designed a new approach to exploit social relations by combining semantic and social signals during the ranking. They put the users, tags, and documents into a friendship graph and applied PageRank computation on it. Similarly, Kashyap et al. \cite{KashyapAH12} designed six social groups and formed a social aware search graph for ranking. Bjorklund et al. \cite{BjorklundGGG11} and Carmel et al. \cite{CarmelZGOHRUYC09} constructed the social network from some social applications, and took the social relations into account while ranking. Recently, some studies concentrated on the personalized search in microblog, which has an explicit Follow-Follower social network. Vosecky et al. \cite{VoseckyLN14} and Zhao et al. \cite{ZhaoLM16} leveraged social relations in Twitter to construct a better user preference for personalized re-ranking of tweets. Different from these social search methods in a specific system, we focus on web search which retrieves the documents from the whole Internet.

\section{Methods}
Referring to the profiles of similar users to personalize the results can improve the ranking quality when the user lacks historical activities. As we stated in Section~\ref{sec:intro}, finding similar users based on simple lexical and topical similarity of query logs is weak in measuring the closeness between users in semantic space. Moreover, when the user has limited interactions with search engine, similar users found based on it are still insufficient to ensure reliability of personalization for new queries. To handle these problems, we propose the model FNPS to highlight similar users in semantic space with neural networks. It integrates the friend network to further address the problem of historical data sparsity. Specifically, to construct the group profile in a fine-grained manner, the user is grouped into multiple friend circles with respect to his search behaviours and friend relations respectively. They are called behaviour-based friend circles and relation-based friend circles in the following. Under the complementary effect of them, similar users are highlighted and a group profile is constructed based on them.

To begin with, suppose that for the user $u$, his historical query log $H_u$ can be divided into the long-term history $H_u^l$ and the short-term history $H_u^s$. The former includes a series of queries and satisfied documents in previous sessions, $H_u^l=\{\{q_1, D_1\},...,\{q_n, D_n\}\}$, where $n$ is the number of queries issued in previous sessions. The latter contains user's recent interactions in the current session, $H_u^s=\{\{q_{n+1}, D_{n+1}\},...,\{q_{t-1}, D_{t-1}\}\}$, where $t$ is the current timestamp. The friend network of the user can be denoted as a graph, $\mathcal{G}=\{\mathcal{V},\mathcal{E}\}$, where $\mathcal{V}$ is the set of nodes containing the current user and his friends, and $\mathcal{E}$ represents the friend relations between users. Given a new query $q$ and candidate documents $D=\{d_1,d_2,...\}$ returned by the search engine, we need to score each element in $D$ to represent the probability of being clicked. The scoring process should refer to the current query $q$, the historical data $H$, and the friend network $\mathcal{G}$. Thus, we denote the score of the document $d$ as $p(d|q,H, \mathcal{G})$, which consists of two parts:
\begin{equation}\label{eq:score}
    p(d|q,H,\mathcal{G})=\phi(p(d|q),p(d|P^I_u,P^G_u,q)),
\end{equation}
where $p(d|q)$ is the adhoc relevance between the document and the query, and $p(d|P^I_u,P^G_u,q)$ represents the personalized relevance based on the user profile. Specifically, the $P^I_u$ denotes $u$'s individual profile, which is constructed with respect to the query $q$ and history $H$. The $P^G_u$ corresponds to $u$'s group profile, which is generated by combining all the information $q$, $H$, and $\mathcal{G}$.
The function $\phi(\cdot)$ is the multilayer perceptron (MLP) with $tanh(\cdot)$ as the activation function, which is used to combine the scores of these two parts with different weights.

\begin{figure*}[!t]
	\centering
	\setlength{\abovecaptionskip}{0.1cm}
	\includegraphics[width=0.93\linewidth, height=8.5cm]{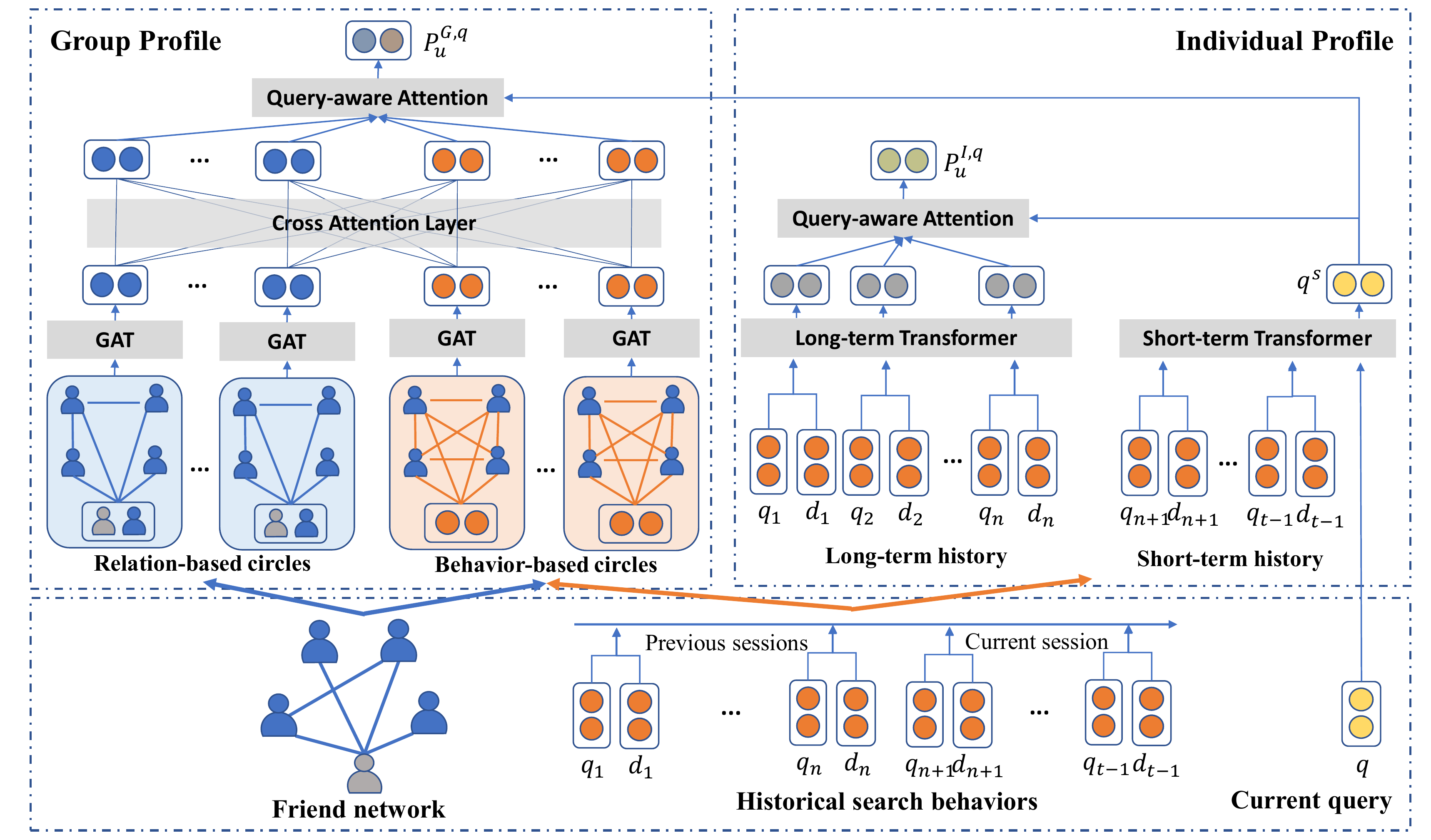}
	\caption{The architecture of FNPS. Given the user's historical search behaviours and current query, the query intent and the individual profile are modeled with transformer layers. With the help of friend network, relation-based and behaviour-based friend circles are formed to build the group profile. These two profiles are integrated to enhance search results personalization.}
	\label{fig:model}
\end{figure*}

The architecture of our model is shown in Figure~\ref{fig:model}. The input consists of the user's friend network, historical search behaviours and the current query. At first, in order to construct the individual profile, we use two transformers to model the long-term and short-term history respectively based on the current query. And then, for building the group profile, we combine the friend relations and search behaviours to form multiple friend circles. Under the interaction between two types of friend circles, more central circles are highlighted to personalize the results. We will introduce the details in the remaining parts of the section.

\subsection{Modeling individual profile}
The key to personalized search is how to model user interests based on user's historical search behaviour. Inspired by previous researches \cite{li2007dynamic, bennett2012modeling, sigir/ZhouDW20}, we model the long-term and short-term historical search behaviour of users separately. The former describes more long-standing user characteristics, and the latter usually represents the user’s recent interests or temporary information needs. Due to the powerful ability of Transformer \cite{vaswani2017attention} on long-term dependency, we attempt to apply it to model the individual profile with the following two steps.

\textbf{Modeling user current query intent.} 
\label{sec:intent}
A common situation in the search engine is that users often put forward a series of queries for a single information need in a session. Therefore, recent interactions in short-term history are helpful for clarifying the intent of current query, especially when the current query is ambiguous. Inspired by the previous work \cite{ZhouDW20}, we apply the short-term transformer to model the current query intent.

For each query and document, the vector representation is computed by summing up the words together based on word embedding, which is trained by the word2vec method \cite{mikolov2013exploiting}. Formally, for each interaction in short-term history, we sum up the query vector and the average satisfied document vector to represent the search intent, denoted as $h_i=q_i+\text{average}(D_i)$. We concatenate the representation of short-term history $H_u^s=\{h_{n+1},...,h_{t-1}\}$ with the current query $q$ as the input of the short-term transformer:
\begin{equation}
\label{eq:qs}
    q^s=\text{Transformer}^{last}([H_u^s,q]+\text{PE}([H_u^s,q])),
\end{equation}
where $\text{Transformer}(\cdot)^{last}$ is implemented by the transformer encoder \cite{vaswani2017attention} and the output of the last position is taken. The function $\text{PE}(\cdot)$ is the position embedding function of transformer. The output $q^s$ indicates the user's current query intent, which contributes to the construction of the user profile and matching with candidate documents in the following.

\textbf{Building individual profile dynamically.}
\label{sec:individual}
Long-term search behaviour often reflects the user's background and stable interest. For example, the user who frequently submits queries related to "pytorch" is more likely to be a programmer. In order to model the user's long-term interests, we devise another transformer to model the long-term dependencies between historical behaviours. Similar to the short-term history, the representations of interactions in long-term history $H_u^l=\{h_1,...,h_n\}$ constitute the input of the long-term transformer:
\begin{equation}
    O_u^l=\text{Transformer}(H_u^l+\text{PE}(H_u^l)),\nonumber
\end{equation}
where $O_u^l=\{o_1,...o_n\}$ is the set of outputs with respect to $u$'s long-term interactions. Intuitively, not all long-term interactions are valuable for current query intent \cite{hrnncikm}. Based on this idea, we dynamically adjust the weight of each interaction $o_i$ with attention mechanism based on the user's current query intent $q^s$. The dynamic individual profile $P_u^{I,q}$ is computed by weighted summation:
\begin{equation}
\label{eq:pi}
   P_u^{I,q}=\sum_{i=1}^n \alpha_i o_i,
\end{equation}
where $\alpha_i$ is the weight of $o_i$ with respect to the current query intent. It is calculated by feeding $o_i$ and $q^s$ into MLP, and is normalized by the softmax function:
\begin{equation}
    \alpha_i=\text{softmax}(\phi(o_i,q^s)),
    \text{softmax}(e_i)=\frac{\text{exp}(e_i)}{\sum_{j=1}^{n}\text{exp}(e_j)}.\nonumber
\end{equation}

Finally, we obtain the individual profile $P_u^{I,q}$ and the current intent $q^s$ related to the current query. They will act on matching with candidate documents and compute the personalized relevance. However, when the user’s historical data is limited, the individual profile is weak for personalization. To tackle this issue, a group profile is built in the following.

\subsection{Friend Circle Formation}
In real life, a user can often be grouped into different friend circles, such as colleagues, relatives, classmates, etc. Each friend circle can reflect an aspect of user characteristics. In this section, we attempt to form multiple friend circles of the user to capture the group profile in a fine-grained way. As we discussed in Section~\ref{sec:intro}, finding similar users based only on search behaviour is not enough when the user lacks historical interactions. The friend relation can also provide us with a way to measure the closeness between users without any behaviours. Based on this consideration, we group the user into multiple friend circles based on relation and behaviour respectively. The former inclines to group users with similar backgrounds into a circle, while the latter focuses on similar information needs. The details of group formation from two angles are as follows.

\begin{algorithm}[t]
\caption{Friend Circle Formation}
\label{alg:circle}
    \begin{algorithmic}
    \REQUIRE friend graph $\mathcal{G}$, candidate nodes $N$, maximum number of friend circle $k$\\
    \ENSURE friend circles $C=\{c_1,...c_k\}$, core nodes $F=\{f_1,...f_k\}$\\
    \FOR{$i$ in range($k$)}
    \FOR{all $n \in N$}
    \STATE E $\leftarrow$ GetNumberofEdges(n) from $\mathcal{G}$
    \ENDFOR
    \STATE $f_i \leftarrow$ N[Argmax(E))]
    \STATE $c_i \leftarrow$ SubGraph($f_i$) from $\mathcal{G}$
    \STATE $\mathcal{G} \leftarrow $ DeleteEdges($c_i$) from $\mathcal{G}$
    \ENDFOR
    \RETURN $C$, $F$
    \end{algorithmic}
\end{algorithm}
\textbf{Relation-based friend circle.}
Since the establishment of friendship in real life is often based on shared experiences, the users in the same relation-based friend circle may reflect the same background. In general, there are some friends who are closer with $u$ in each circle. Forming friend circles based on these close friends can more accurately reflect the user's group information. To find the close friends, we take the number of mutual friends as the indicator to measure the closeness between users.

Formally, we develop an algorithm of friend circle formation as described in Algorithm~\ref{alg:circle}. Given the user's friend network $\mathcal{G}=\{\mathcal{V},\mathcal{E}\}$ and the candidate nodes $N$, our goal is to find out the close friends from $N$ and to form top $k_\text{r}$ friend circles based on them. The candidate nodes here contain all friends of the current user $u$, $N=\{v_i|v_i \in \mathcal{V},v_i \neq u\}$. Specifically, we first select the node with the most edges on $\mathcal{G}$ from candidate nodes $N$, which represents the friend who has the most mutual friends with user $u$. The subgraph related to this node on $\mathcal{G}$ is regarded as the first friend circle $c_1$. The node can be regarded as the core of this friend circle $f_1$, which has the same friend relations as the user $u$ in $c_1$. To ensure the distinction between different circles, we remove the edges of this friend circle from $G$ and repeat the above process to form other circles. Finally, we obtain $k_\text{r}$ diverse relation-based friend circles $C^\text{r}=\{c^\text{r}_1,...c^\text{r}_{k_\text{r}}\}$ and corresponding core friends $F^\text{r}=\{f^\text{r}_1,...f^\text{r}_{k_\text{r}}\}$. They will act on modeling the group profile in the next section.

\textbf{Behaviour-based friend circle.}
The formation of friend circle is not only based on the same background, but may also establish upon similar interests, such as a sport, a movie star, etc. Historical search behaviour can reflect user interests to a certain extent. In this part, we attempt to group users based on their historical search behaviours. 
To implement this idea, the edges of the friend network $\mathcal{G}$ are replaced based on the common search behaviour in this section. The users who issued the same query or were satisfied with the same documents are connected with an edge. Moreover, we put $u$'s historical search behaviours (queries and satisfied documents) into the graph to form a heterogeneous graph. Each query or document is regarded as a node and the other users whose query logs contain it are connected with this node. Formally, this new heterogeneous graph is considered as the behaviour-based friend graph, and the nodes which represent $u$'s search behaviours are set as candidate nodes. We apply the same algorithm as above on them to group the user into different friend circles based on search behaviours. Finally, we generate top $k_\text{b}$ behaviour-based friend circles $C^\text{b}=\{c^\text{b}_1,...c^\text{b}_{k_\text{b}}\}$ and the corresponding core behaviours. Note that since the core behaviour must appear in the query log of each friend in the circle, each behaviour-based friend circle is a fully connected graph. 

These two types of friend circle focus on different angles of grouping. We believe they complement each other and can build a more comprehensive group profile. The following section will introduce how to integrate them to get the group profile.

\subsection{Constructing Group Profile}
Although we formed multiple friend circles according to the friend relation and search behaviour, how to aggregate these information is still a challenge. In this section, we propose three insights on modeling group profile. Firstly, we hold the idea that the users in the friend circle have different distances to the current user. Secondly, relation-based and behaviour-based friend circles are formed from two angles. They complement each other and the users who are essential in both dimensions should be given more attention. Thirdly, the contribution of each circle is different in facing different information needs. We will elaborate on the role of each opinion and the implementation details in the following.

\textbf{GAT on friend circle.}
The friend circle we formed in last section can be regarded as a group of users with similar background or interest around the core node. Based on the first insight, we intend to explore the structure information of the circle to characterize the influence of each friend. Due to the powerful ability of Graph Attention Network (GAT) \cite{VelickovicCCRLB18} in modeling the structure of the graph, we apply it on each friend circle to highlight similar users.

Specifically, we model the user profile of each friend based on his historical behaviours referring to Section~\ref{sec:individual}. For the group user $g$ in friend circles, we apply the long-term transformer with the same parameters and average the outputs to represent his individual profile $P^I_g$. Formally, for each friend circle $c_i$ and the representation of core node $f_i$, the aggregation function of GAT is defined as:
\begin{equation}
    c^f_i= \text{LeakyReLU}(W\sum_{g \in c_i} \alpha_{ig} P^I_g), \alpha_{ig}=\text{softmax}(\phi(f_i,P^I_g)).\nonumber\nonumber
\end{equation}
where $c^f_i$ is regarded as the feature of the circle $c_i$. The $W$ is the model parameter which is learned during the training. The $\alpha_{ig}$ is the aggregation weight of the user $g$ based on the core node and it is parameterized with MLP. The representations of relation-based friend circles $C^{\text{r},f}$ and behaviour-based friend circles $C^{\text{b},f}$ will interact in the next step to complement each other.

\textbf{Cross attention on relation and behaviour.}
As we proposed above, the relation-based friend circles tend to dig out the background information, while the behaviour-based friend circles capture the interests of the user. Based on the second insight, we believe the friends who appear in both types of circles have more contribution to the user profile. In other words, if a relationship-based friend circle and a behaviour-based friend circle contain many common users, we should pay more attention to these users through the interaction of these two circles.

To achieve such an interaction between the two types of friend circles, we propose to use a masked transformer which only keeps the connections between different types of circles. We concatenate the representations of two types of circle $C^{\text{r},f}$ and $C^{\text{b},f}$, and feed them into the masked transformer. We have:
\begin{equation}
    C^f = \text{Transformer}^{masked}([C^{\text{r},f}, C^{\text{b},f}]),\nonumber
\end{equation}
where $\text{Transformer}^{masked}(\cdot)$ means adding the mask matrix $M$ to the weight matrix of the transformer encoder before softmax. The computation of the mask matrix $M$ is based on the relations among different friend circles:
\begin{align*}
\begin{split}
M_{ij}= \left \{
\begin{array}{ll}
   0 & \text{if $c_i^f$ and $c_j^f$ are connected}\\
   -inf & \text{if $c_i^f$ and $c_j^f$ are disconnected}
\end{array}
\right.
\end{split}
\end{align*}
The output of cross-attention layer $C^f=\{c^f_1,...c^f_{k_\text{r}+k_\text{b}}\}$ represents the enhanced friend circles combining the information of search behaviour and friend relation. It provides a foundation for building a dynamic group profile in the next step.

\textbf{Query-aware attention on different circles.}
\label{sec:friend}
Intuitively, not all friend circles are helpful when the user issues a new query. For adjusting the weight of each friend circle, we apply an attention mechanism on different circles with respect to the user's current query intent. Formally, we take the output of the cross-attention layer $C^f$ and the user's query intent $q^s$ to learn the weight of each circle by MLP. Similar to the Eq.~(\ref{eq:pi}), we have:
\begin{equation}
\label{eq:pf}
    P_u^{G,q} = \sum_{i=1}^{k_\text{r}+k_\text{b}} \alpha_i c^f_i, \alpha_i = \text{softmax}(\phi([c^f_i, q^s])).
\end{equation}
The final output $P_u^{G,q}$ is the group profile, which is constructed based on friend network, historical search behaviours, and the current query. It will play an essential role in the search results personalization, especially when the individual profile is weak in personalizing the results of current query.

\subsection{Search Results Personalization}
In this section, we compute the score of each candidate document based on the individual profile and the group profile. The computation of each part in Eq.~(\ref{eq:score}) is introduced as follows.

For adhoc relevance $p(d|q)$, we take the vector similarity between the query and the document to represent their semantic relevance. Moreover, we also extract a series of relevance features $\mathcal{F}_{q,d}$ following \cite{bennett2012modeling}, mainly including original ranking, click features, and topical features. We use MLP to aggregate these features to compute a relevant score. The adhoc relevance consists of two parts:
\begin{equation}
    p(d,q)=\phi\left(\text{sim}(q,d),\phi(\mathcal{F}_{q,d})\right),\nonumber
\end{equation}
where the function $\text{sim}(\cdot)$ is computed by cosine similarity. 

The personalized relevance $p(d|P^I_u,P^G_u, q)$ is related to the individual profile and the group profile we obtained above. Intuitively, the group profile is more useful when the individual profile cannot provide effective personalized information for the current query. Therefore, we devise a gate unit to combine two parts with regard to the current query $q$ and individual profile $P^{I,q}_u$ from Eq.~(\ref{eq:pi}). The gate weight is computed by feeding them into MLP, denoted as $c=\phi([q,P^{I,q}_u])$. Finally, the personalized relevance combines the individual part and group part with this weight:
\begin{equation}
    p(d|P^I_u,P^G_u,q)=c*p(d|P^I_u,q) + (1-c)*p(d|P^G_u,q). \nonumber
\end{equation}
For the individual part, the current query intent $q^s$ from Eq.~(\ref{eq:qs}) and the individual profile $P^{I,q}_u$ are matched to the candidate documents. And the matching between the profile $P^{G,q}_u$ from Eq.~(\ref{eq:pf}) and the document $d$ represents the score of group part. We have:
\begin{equation}
\begin{split}
    p(d|P^I_u,q) &= \phi\left(\text{sim}(d,P^{I,q}_u),\text{sim}(d,q^s)\right), \\
    p(d|P^G_u,q) &= \text{sim}(d,P^{G,q}_u).\nonumber
\end{split}
\end{equation}

Finally, the results are personalized by re-ranking the document list according to the score of each candidate documents. We adopt the LambdaRank algorithm \cite{wu2008ranking} to train our ranking model, which constructs the document pair with a satisfied document and an irrelevant document. Formally, the distance between the positive sample $d_i$ and the negative sample $d_j$ is computed by $|p(d_i|q,H,\mathcal{G})-p(d_j|q,H,\mathcal{G})|$ with sigmoid normalization, denoted as $p_{ij}$. The loss function is defined as the weighted cross entropy between true distance and predicted distance:
\begin{equation}
    \mathcal{L}=-|\Delta|\left(\overline{p}_{ij}log(p_{ij})+\overline{p}_{ji}log(p_{ji})\right),\nonumber
\end{equation}
where the weight $\Delta$ represents the change of ranking quality after swapping the position of $d_i$ and $d_j$. By minimizing the loss function with Adam optimizer, the model gradually reaches convergence.
\begin{table}[!t]
  \caption{Basic statistics of the dataset.}
  \label{tab:querylogs}
  \setlength{\abovecaptionskip}{0.1cm}
  \setlength{\belowcaptionskip}{0.1cm}
  \begin{tabular}{cc||cc}
  	\toprule
    Item & Statistic & Item & Statistic \\
  	\midrule
  	\#days & 92 & \#seed users & 43,770 \\
    \#friends & 2,195,625 & \#friend relations & 126,860,976 \\
  	\#queries & 14,930,839 & \#distinct queries & 9,334,434 \\
	\#clicks & 15,843,926  & \#sessions & 8,226,296\\
  	\bottomrule
  \end{tabular}
\end{table}

\begin{table*}[!t]
 \center
 \setlength{\abovecaptionskip}{0.1cm}
 \setlength{\belowcaptionskip}{0.1cm}
 \caption{Overall performance of FNPS and other baselines. The percentage reflects improvements over original ranking. "$\dagger$" indicates the model outperforms all baselines significantly with paired t-test at p $<$ 0.05 level. Best results are shown in bold.}
  \label{tab:aol}
  \begin{tabular}{p{0.1\linewidth}|p{0.027\textwidth}l|p{0.027\textwidth}l|p{0.027\textwidth}l|p{0.027\textwidth}l|p{0.027\textwidth}l|p{0.027\textwidth}l|p{0.027\textwidth}l}
  	\toprule
  	Model &\multicolumn{2}{c|}{MAP} & \multicolumn{2}{c|}{MRR} & \multicolumn{2}{c|}{P@1} &\multicolumn{2}{c|}{Ave. Rank} & \multicolumn{2}{c|}{NDCG@3} & \multicolumn{2}{c|}{NDCG@5} & \multicolumn{2}{c}{NDCG@10} \\ \hline
  	Ori. & .6614 & - & .6859 & -  & .5766 & - & 3.898 & - & .6330 & - & .6720 & - & .7148 & -\\ \hline
  	\multicolumn{11}{l}{Group based methods} \\ \hline
  	G-Click & .6625 & +0.17\% & .6874 & +0.22\%  & .5746 & -0.35\% & 3.849 & +1.26\% & .6371 & +0.65\% & .6754 & +0.51\% & .7166 & +2.52\% \\
  	DGF & .6491 & -2.02\% & .6769 & -1.31\%  & .5583 & -3.17\% & 3.920 & -0.56\% & .6253 & -1.22\% & .6616 & -1.55\% & .7069 & -1.11\% \\ \hline
  	  
  	\multicolumn{11}{l}{Social network based methods} \\ \hline
  	PSSN & .6642 & +0.42\% & .6883 & +0.35\% & .5732 & -0.59\% & 3.812 & +2.21\% & .6389 & +0.93\% & .6778 & +0.86\% & .7193 & +0.63\%\\
  	SonetRank & .6659 & +0.68\% & .6898 & +0.57\% & .5731 & -0.61\% & 3.764 & +3.44\% & .6409 & +1.25\% & .6796 & +1.13\% & .7220 & +1.01\%\\ \hline
  	\multicolumn{11}{l}{Deep learning based methods} \\ \hline
	HRNN & .6707 & +1.41\% & .6951 & +1.34\% & .5800 & +0.59\% & 3.727 & +4.39\% & .6478 & +2.34\% & .6839 & +1.77\% & .7258 & +1.54\%\\
	RPMN & .6724 & +1.66\% & .6962 & +1.50\% & .5795 & +0.50\% & 3.642 & +6.57\% & .6475 & +2.29\% & .6859 & +2.07\% & .7296 & +2.07\%\\
	PEPS & .6727 & +1.71\% & .6979 & +1.75\% & .5811 & +0.78\% & 3.626 & +6.98\% & .6501 & +2.70\% & .6867 & +2.19\% & .7303 & +2.17\%\\ 
	HTPS & .6749 & +2.04\% & .6991 & +1.92\% & .5821 & +0.95\% & 3.615 & +7.26\% & .6513 & +2.89\% & .6894 & +2.59\% & .7326 & +2.49\%\\ \hline
	\multicolumn{11}{l}{Our designed model} \\ \hline
    FNPS & $\bm{.6827}^\dagger$ & +3.22\% & $\bm{.7058}^\dagger$ & +2.90\% & $\bm{.5902}^\dagger$ & +2.36\% & $\bm{3.486}^\dagger$ & +10.57\% & $\bm{.6585}^\dagger$ & +4.03\% & $\bm{.6985}^\dagger$ & +3.91\% & $\bm{.7402}^\dagger$ & +3.55\%\\
    \bottomrule
  \end{tabular}
\end{table*}
\section{Experimental Setup}
\subsection{Dataset and Evaluation Metrics}
There is no public dataset that have both search logs and friend networks with shared users. To evaluate the effectiveness of the model, we collect the data from a large social platform, which embeds a search engine for users to acquire information. Since the social network of the platform is too large for experiments, we extract a small friend network from it for experimentation. To avoid too sparse friend network, we randomly sample some seed users and take their friends to form the friend network in experiments. We collect their search behaviours in the last three months of 2020. Each piece of data in the query logs contains \textit{an anonymous user ID, a query string, query issued time, top 20 documents returned by search engine, clicked documents, and click dwell times}. Note that the returned results include not only the contents created in the platform, but also the general web pages of the entire Internet. The basic statistics are shown in Table~\ref{tab:querylogs}.

Since personalization requires historical behaviours to build basic user profile, we regard the first 8 weeks data as the history. The last 5 weeks data is divided with 4:1:1 ratio for training, validation, and testing. To ensure that there is at least one session for each part, we remove the users whose active time is less than four sessions. The session is demarcated by 30 minutes of user inactivity \cite{white2007investigating}.

To evaluate the ranking quality, we define the satisfied documents as the documents whose click dwell time is more than 30s \cite{bennett2012modeling, vu2015temporal}. Moreover, we also identify a document as relevant if it is satisfied under the next two queries following \cite{hrnncikm, bennett2012modeling}. We choose mean average precision (MAP), mean reciprocal rank (MRR), precision@1 (P@1), average click position (Avg. Click), and normalized discounted cumulative gain@K (NDCG@K) to measure the results.

\subsection{Baselines}
The original ranking of the search engine is considered as a basic baseline. Additionally, we compare our model with the group based methods, social network based methods, and deep learning based methods for personalized search. 

\textbf{G-Click} \cite{dou2007large}. This is a method of grouping based on click behaviours, which scores documents by counting the same behaviours of other users, thereby re-ranking the list.

\textbf{DGF} \cite{Vu2014Improving}. It finds similar users based on topic features constructed using the LDA method, and then considers the interests of these similar users when personalizing search results.

\textbf{PSSN} \cite{CarmelZGOHRUYC09}. It achieves personalization based on the user’s social relations, which considers the familiarity-based network, similarity-based network, and overall network. We regard the friend network as the familiarity-based network in this baseline.

\textbf{SonetRank} \cite{KashyapAH12}. It builds a Social Aware Search Graph, which consists of groups, users, queries and clicked documents. It aggregates the relevance feedback of the similar users in the group. We treat the user's friends as a group of similar users.

\textbf{HRNN} \cite{hrnncikm}. This is a deep learning based method which leverages the sequential information hidden in query logs by RNN. Attention mechanism is applied to build dynamic user profiles.

\textbf{RPMN} \cite{ZhouDW20}. It enhances re-finding behaviours in personalized search with memory networks. Query-based re-finding and document-based re-finding are considered for different search intent.

\textbf{PEPS} \cite{sigir/YaoDW20}. It trains the personal embedding matrix for each user based on historical behaviour. The global embedding and personal embedding are taken into account in personalization.

\textbf{HTPS} \cite{sigir/ZhouDW20}. It abandons the construction of user profiles, and instead encodes the history as context to disambiguate the query with Transformer encoder, which is also used in our model.

The first two models group users based on their click information, while the next two methods leverage the social network to find similar users. The last four approaches mine user preferences hidden in historical behaviours with deep learning. Our method combines the advantages of these baselines and we named the model as FNPS (\textbf{F}riend \textbf{N}etwork enhanced \textbf{P}ersonalized \textbf{S}earch)\footnote{The code of the model is available on https://github.com/smallporridge/FNPS}.

The parameters of the model are selected through multiple experiments. We choose the word embedding size in $\{\bm{100}, 300\}$, the hidden state size of transformer and GAT in $\{128, 256, \bm{512}\}$, the number of MLP hidden units in $\{128, \bm{256}, 512\}$, the learning rates in $\{10^{-2}, \bm{10^{-3}}, 10^{-4}\}$. For hyperparameters $k_\text{r}$ and $k_\text{b}$, we compute them according to the number of friends and search behaviours, $k_\text{r}=\frac{\text{\#friends}}{20}$, $k_\text{b}=\frac{\text{\#behaviours}}{5}$. Considering the performance and memory usage, we select the parameters in bold to train the model. 

\begin{table*}[!t]
 \center
 \setlength{\abovecaptionskip}{0.1cm}
 \setlength{\belowcaptionskip}{0.1cm}
 \caption{Performance of ablation studies of the FNPS model. The percentage is calculated based on the whole model.}
  \label{tab:ablation}
  \begin{tabular}{p{0.1\linewidth}|p{0.027\textwidth}l|p{0.027\textwidth}l|p{0.027\textwidth}l|p{0.027\textwidth}l|p{0.027\textwidth}l|p{0.027\textwidth}l|p{0.027\textwidth}l}
  	\toprule
  	Model &\multicolumn{2}{c|}{MAP} & \multicolumn{2}{c|}{MRR} & \multicolumn{2}{c|}{P@1} &\multicolumn{2}{c|}{Ave. Rank} & \multicolumn{2}{c|}{NDCG@3} & \multicolumn{2}{c|}{NDCG@5} & \multicolumn{2}{c}{NDCG@10} \\ \hline
  	\;\;w/o. RGF& .6766 & -0.89\% & .7001 & -0.81\% & .5838 & -1.25\% & 3.611 & -1.08\% & .6526 & -0.90\% & .6924 & -0.87\% & .7349 & -0.72\%\\
  	\;\;w/o. BGF & .6772 & -0.80\% & .7008 & -0.72\% & .5840 & -1.22\% & 3.604 & -1.04\% & .6531 & -0.82\% & .6930 & -0.78\% & .7356 & -0.62\%\\
  	\;\;w/o. GAT & .6790 & -0.54\% & .7023 & -0.50\% & .5870 & -0.37\% & 3.585 & -0.54\% & .6549 & -0.55\% & .6950 & -0.49\% & .7373 & -0.39\%\\
  	\;\;w/o. CA & .6801 & -0.37\% & .7034 & -0.34\% & .5876 & -0.27\% & 3.570 & -0.44\% & .6562 & -0.35\% & .6964 & -0.29\% & .7383 & -0.26\%\\
	\;\;w/o. QA & .6793 & -0.49\% & .7026 & -0.45\% & .5869 & -0.39\% & 3.579 & -0.69\% & .6551 & -0.52\% & .6953 & -0.45\% & .7374 & -0.38\%\\ \hline
   FNPS & .6827 & - & .7058 & - & .5902 & - & 3.486 & - & .6585 & - & .6985 & - & .7402 & -\\
    \bottomrule
  \end{tabular}
\end{table*}
\section{Results and Analysis}
\subsection{Overall Performance and Analysis}
The overall results of models are listed in Table~\ref{tab:aol}. Some findings are summarized as follows:

(1) Our proposed model FNPS outperforms all baselines on all evaluation metrics, which shows the effectiveness of our friend network enhanced personalized search model on constructing a more reliable user profile. Compared with the best personalized baseline model HTPS, our model improves the ranking quality significantly with paired t-test at p $<$ 0.05 level. Specifically, FNPS expands the improvement on MAP over original ranking from 2.04\% to 3.22\%. A more noteworthy evaluation metric is P@1, on which the improvement increases from 0.95\% to 2.36\%. These results confirm that the behaviour of similar users can provide effective personalized signals when re-ranking the results. And our model is able to accurately strengthen the influence of similar users by integrating search behaviour and friend network.

(2) Comparing different types of personalization methods, we find that group based methods have minimal improvement in results and even have side effects. The DGF model, which uses topical similarity to find similar users, causes a serious decline on the results. A possible reason is that in the case where the quality of the original ranking is high, users with similar topics are not enough to distinguish the candidate documents. Social network based methods achieve better results by introducing social relations to find similar users, which indicates the familiar users tend to have similar interests. Moreover, deep learning based methods show powerful ability to learn user profiles based on historical interactions automatically, and they outperform previous models significantly. Our model combines the advantages of these methods and further introduces the friend network into personalized search with deep learning, which is proven to be effective.

(3) The improvement on P@1 is obviously lower than other metrics, which may be caused by characteristics of mobile search. In this scenario, the user’s re-finding behaviour may be weaker than searching on the computer, such as searching for "Gmail" every day to check emails. Users sometimes want to get latest content instead of what they have seen before. Therefore, it may not be the best choice to rank the viewed documents at the first position. The group based methods and social network based methods even get worse results on P@1. A possible reason is that these methods do not take timeliness into account, which is a key feature in mobile search. All of deep learning based methods extract the sequential information of query logs and achieve better results. Our model has a significant advantage on P@1, which verifies the reliability of re-ranking the results according to the profiles of similar users.

In summary, \textbf{our model integrates search behaviour and friend network to improve the reliability of finding similar users, and a more accurate group profile can be constructed based on them with neural networks for personalized search.}
\begin{figure}[!t]
	\centering
	\vspace{-0.5cm}
	\setlength{\abovecaptionskip}{0.1cm}
	\includegraphics[width=0.65\linewidth,height=0.4\linewidth]{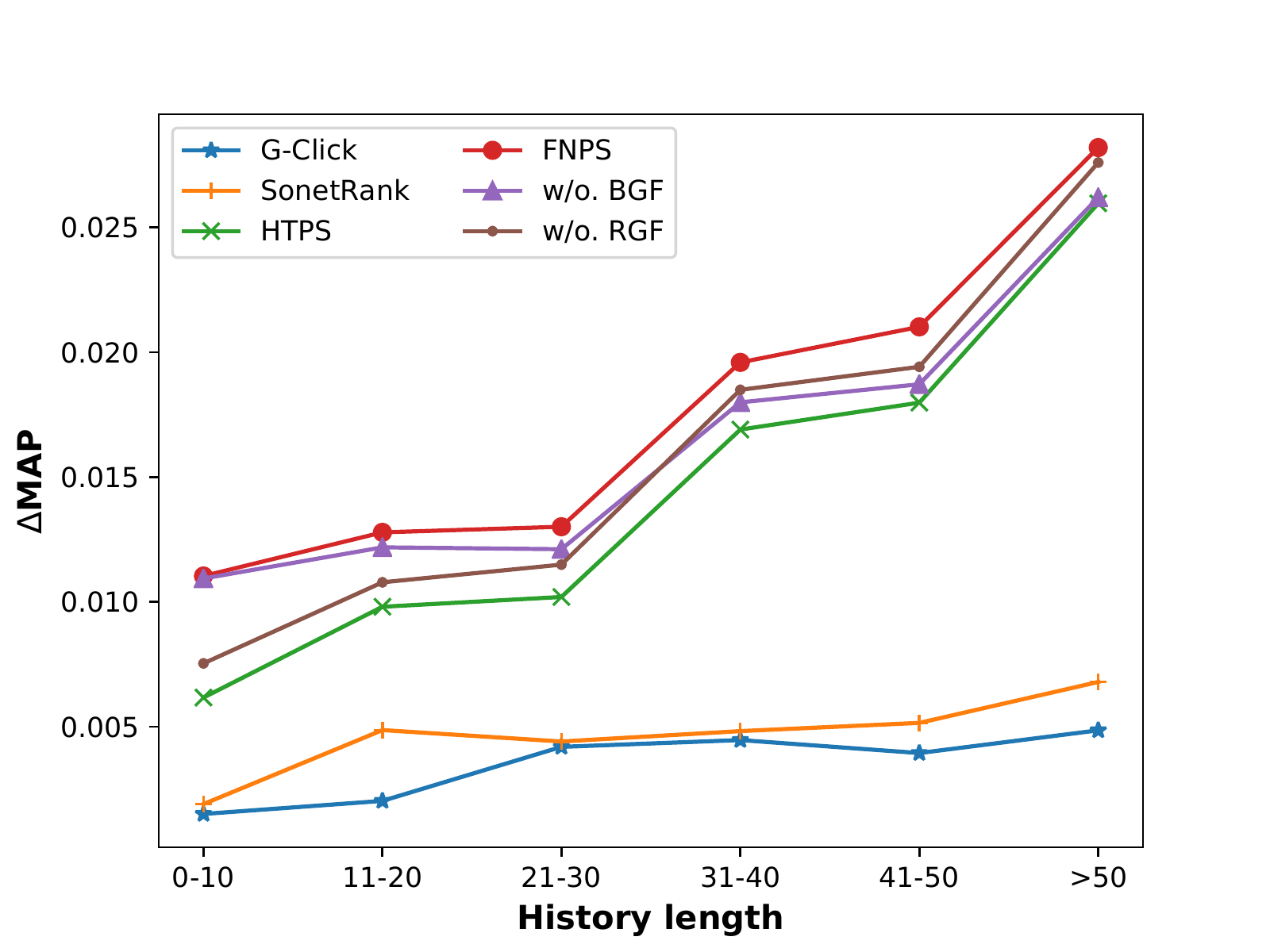}
	\caption{The study of different lengths of history.}
	\label{fig:hislen}
\end{figure}
\subsection{Ablation Experiments}
The FNPS model mainly proposes a group formation method combining the friend relation and search behaviour, and then uses neural networks to model the group profile for personalized search. To verify the effectiveness of each component in this process, we conduct ablation experiments on the relation-based group formation, the behaviour-based group formation, the GAT, the cross attention layer, and the query-aware attention. Specifically, the implementation details are as follows.


\textbf{FNPS w/o. RGF.} We remove the relation-based group formation and only keep the behaviour-based friend circles.

\textbf{FNPS w/o. BGF.} The behaviour-based group formation is removed and only the relation-based friend circles are reserved.

\textbf{FNPS w/o. GAT.} We replace the GAT with averaging the vector of users in the friend circle to represent this circle.

\textbf{FNPS w/o. CA.} We discard the cross attention layer, which means there is no interaction between two types of friend circles.

\textbf{FNPS w/o. QA.} We abandon the query-aware attention when building the group profile. Instead, we simply average the representation of each circle after cross attention layer.

The ablation results are shown in Table~\ref{tab:ablation}. It can be observed that all ablation models perform worse than the whole framework. Specifically, removal of either relation-based or behaviour-based group formation causes considerable damage to the results, which shows the necessity of each type of friend circle on building the group profile. The contribution of the other three components are relatively smaller, but they still have a certain impact on the results. The drop caused by removing GAT shows that deep learning can discover common interests in a friend circle, thereby reducing the impact of irrelevant behaviours of others. Eliminating cross attention layer leads to about 0.37\% decline on MAP, which reveals that two types of friend circles are complementary in building the group profile. The effectiveness of query-aware attention has been proved by the previous work [11]. Similarly, our results consistently show that discarding the query-aware attention will reduce the accuracy of the user profile and damage the results.

\begin{figure}[!t]
	\centering
	\vspace{-0.5cm}
	\setlength{\abovecaptionskip}{0.1cm}
	\includegraphics[width=0.65\linewidth,height=0.4\linewidth]{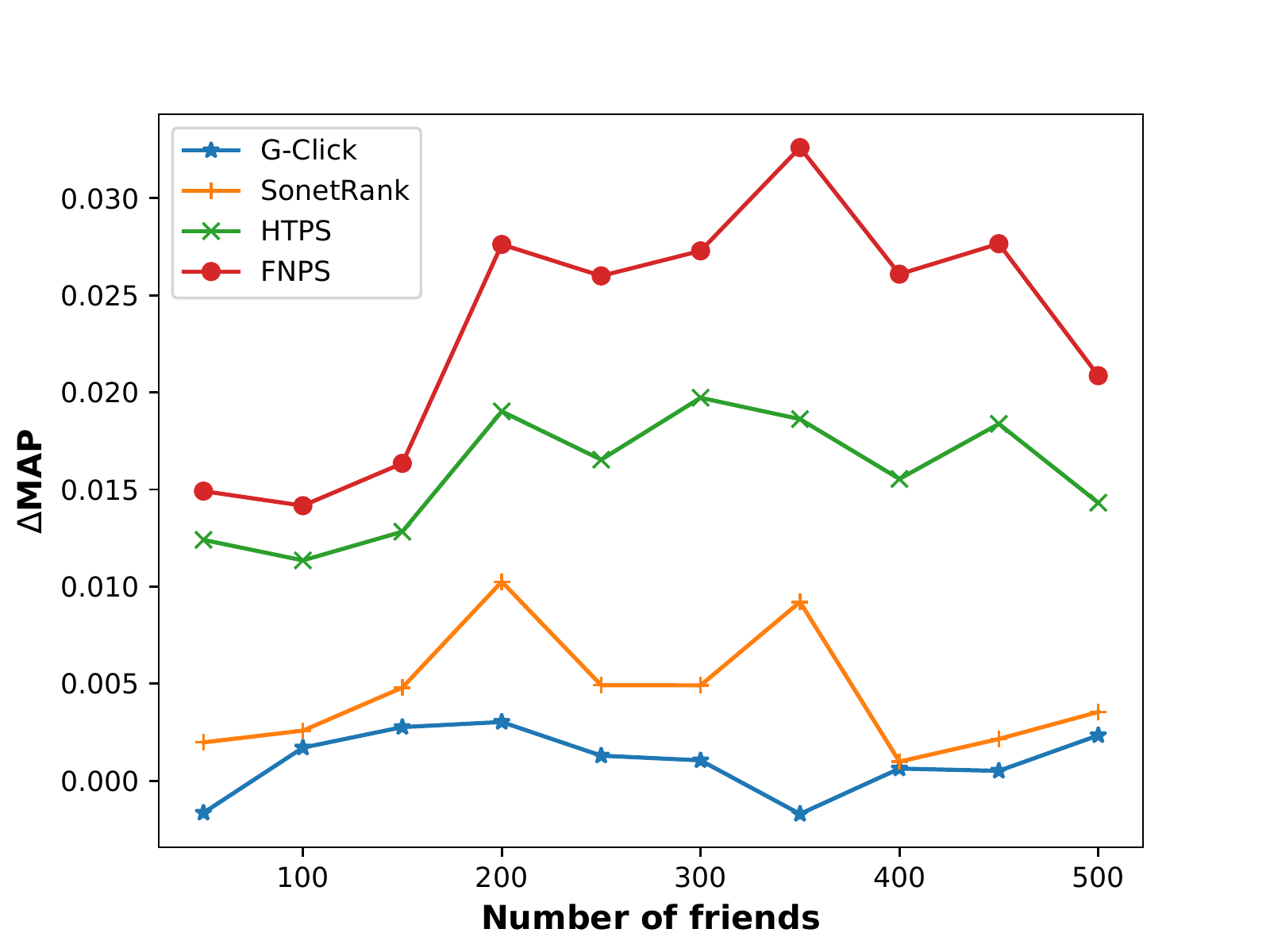}
	\caption{The study of different numbers of friends.}
	\label{fig:friendnum}
\end{figure}
\subsection{Comparison of Different History Lengths}
Personalized search usually depends on the user's historical behaviour. The more the user interacts with the search engine, the more accurate the user profile is constructed by the model. In order to analyze the impact of different history lengths on the model more specifically, we divide the queries with different history lengths into different groups at 10 intervals. We choose G-Click, SonetRank, and HTPS models as the representatives of each type of baseline methods. To observe the role of the two kinds of friend circles, we also take w/o. RGF and w/o. RGF for comparison. Here we calculate the MAP improvement over original ranking to show the performance.

The results are shown in Figure~\ref{fig:hislen}. It can be seen that a longer history helps all personalization methods to improve the ranking quality in general. As the history length grows, the deep learning based methods show more powerful ability to extract the personalized information in the history. Comparing our model FNPS and the best baseline model HTPS, we find that when the user’s historical activity is limited, our model improves more obviously. As the history length increases, the gap between them becomes smaller. This indicates that \textbf{when the user has enough search history, the user’s query intent can be well determined according to his individual profile. The group profile is more useful when the individual profile is weak in personalizing the results}. The performance of the two ablation models demonstrates that the relation-based and behaviour-based friend circles are complementary. Removing RGF causes more drops when the history length is short, while discarding BGF damages the results obviously for queries with long history. Two grouping ways focus on different scenarios and the combination of them makes our model more robust. 

\subsection{Effect of the Number of Friends}
Since our personalization model is based on a friend network, the number of friends will have an impact on the model effect. To explore this issue, we divide users into multiple groups according to the number of friends at intervals of 50, and test the performance of different models. For comparison, we select two friend network independent baselines G-Click and HTPS, and two relation-based models SonetRank and FNPS to conduct the experiment.

The results shown in Figure~\ref{fig:friendnum} reveal that our model FNPS outperforms all baselines in each interval. We also find that friend network related models fluctuate more when the number of friends changes, while the G-Click and HTPS models are relatively stable. Comparing our model FNPS with the best baseline model HTPS, the gap between them implies the contribution of introducing friend network into personalized search for better group formation. It can be seen that the contribution of the friend network is growing when the number of friends increases from 0 to 350. This demonstrates that \textbf{more friends can provide more effective reference information for the model to optimize the user profile}. However, when the number of friends is larger than 350, the gap between HTPS and FNPS is getting smaller. This indicates that too many friends will lead to more noise in modeling the group profile.

\begin{figure}[!t]
	\centering
	\vspace{-0.2cm}
	\setlength{\abovecaptionskip}{0.1cm}
	\includegraphics[width=0.65\linewidth,height=0.4\linewidth]{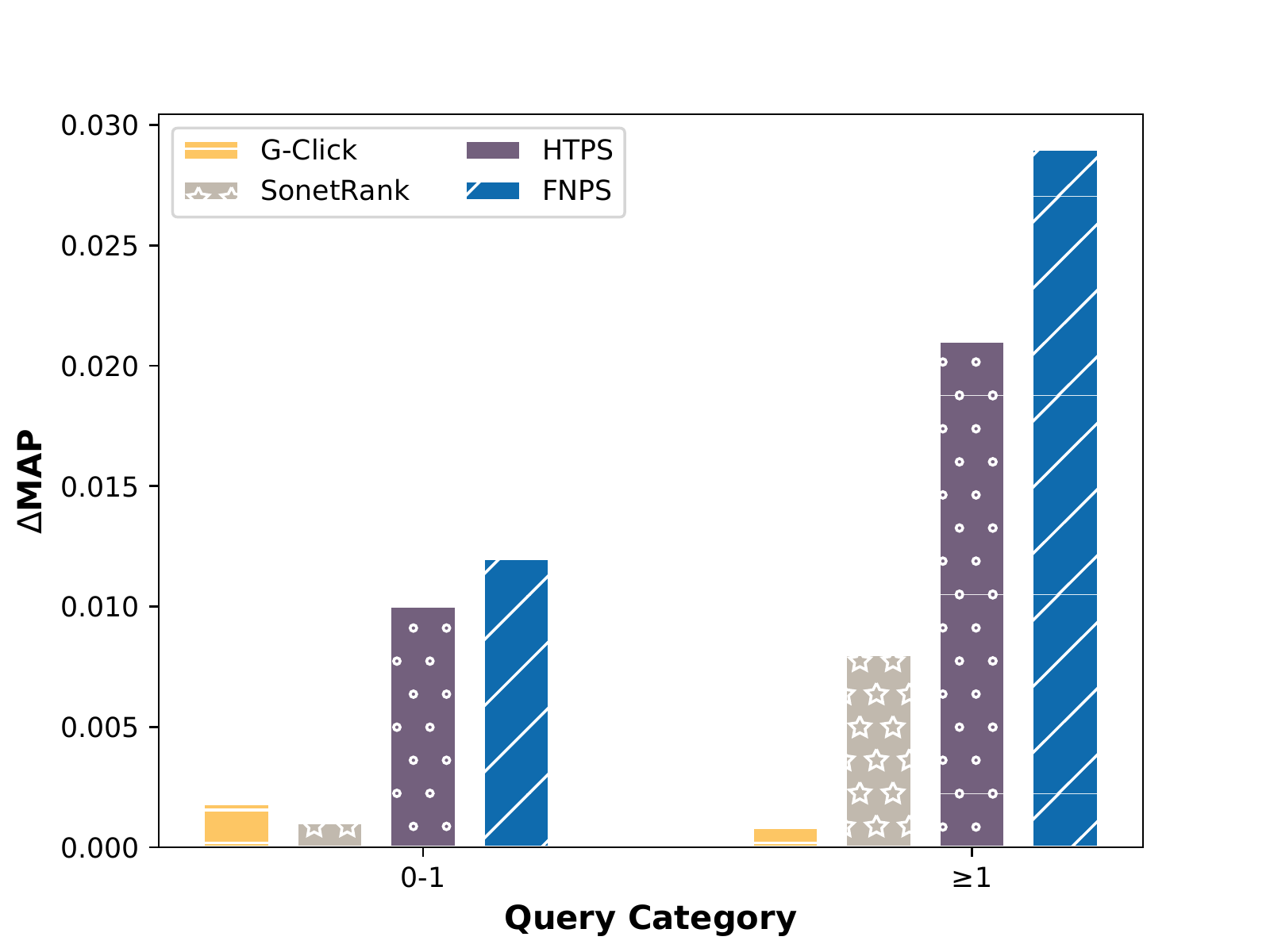}
	\caption{The study of queries with different click entropies.}
	\label{fig:entropy}
\end{figure}
\subsection{Effect of Different Click Entropy}
In search engines, the query can be divided into navigational query and non-navigational query according to the purpose of the query. The intent of the former is clearer and has little diversity for different users, while the latter often has multiple meanings. We choose the click entropy \cite{dou2007large} to measure the ambiguity of the query and divide the tested queries with cutoff of click entropy at 1.0. Queries with larger click entropy are ambiguous which require more personalization. To study the contribution of our model, we compare the improvement over original ranking of FNPS with the baselines models G-Click, SonetRank, and HTPS on the two query sets.

As shown in Figure~\ref{fig:entropy}, all personalized models outperform the original ranking on both query sets, while the improvement on non-navigational queries is more obvious. Specifically, deep learning based models show significant improvement over the traditional models. Compared to the best baseline model HTPS, our model improves the ranking quality on both query sets, especially on the query set with larger click entropy. This indicates that by integrating search behaviour and friend network, a more accurate group profile can be constructed to clarify the query intent.

\begin{figure}[!t]
	\centering
	\vspace{-0.2cm}
	\setlength{\abovecaptionskip}{0.1cm}
	\includegraphics[width=0.65\linewidth,height=0.4\linewidth]{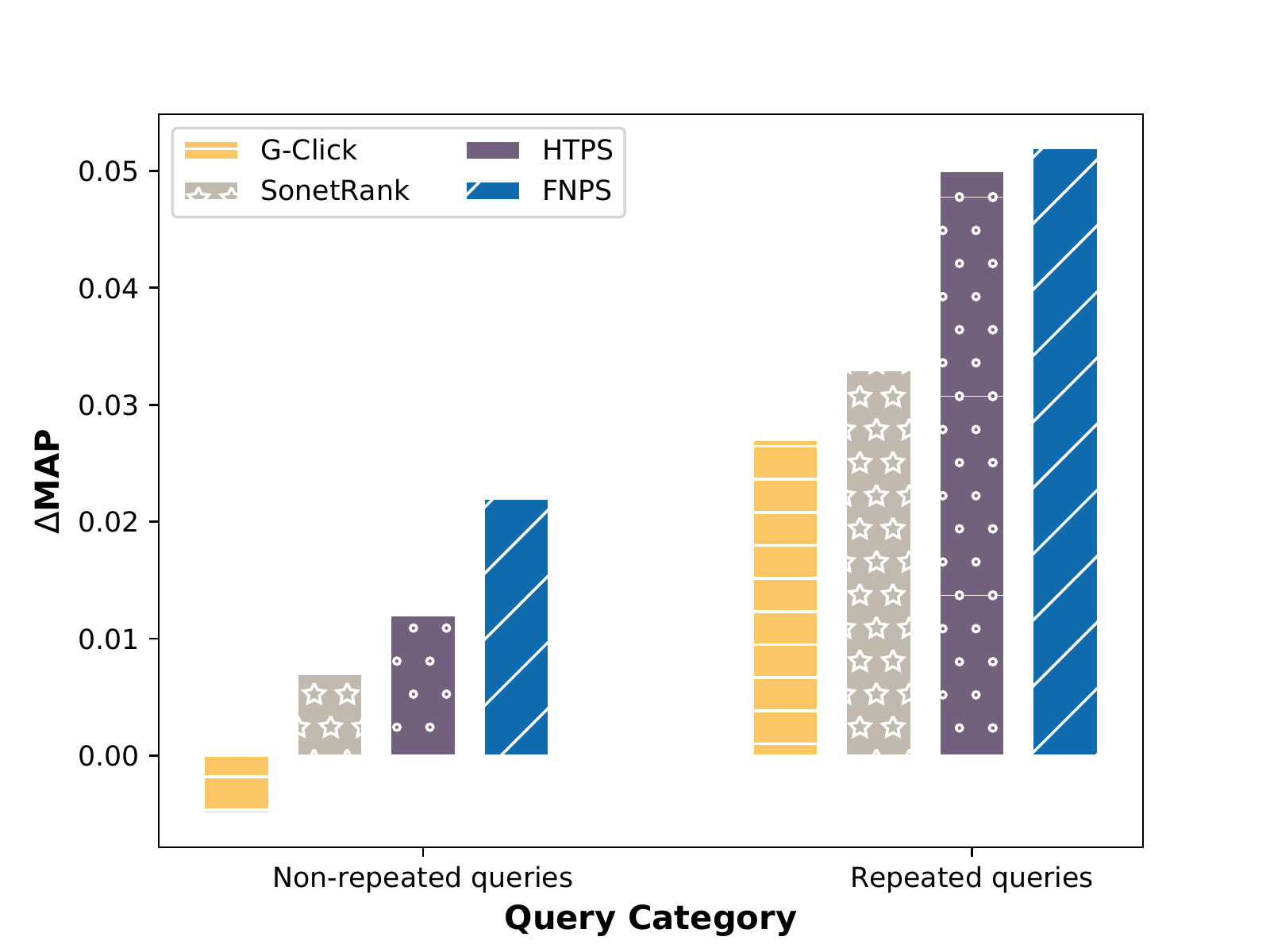}
	\caption{The study of repeated and non-repeated queries.}
	\label{fig:repeat}
\end{figure}
\subsection{Repeated Queries vs. Non-repeated Queries}
In this experiment, we categorize the test query sets into repeated and non-repeated queries. A repeated query is a query that the user has submitted before, which is easy to infer the query intent based on the historical click-through data. But for the non-repeated queries, it is hard to understand the user's intent directly based on existing click-based features. According to our statistics, about 75\% of queries are issued for the first time. The understanding of these queries will greatly contribute to the improvement of the results. To compare the performance of different models on these two query sets, we choose the same baselines as the previous experiment.

From Figure~\ref{fig:repeat}, we observe that all personalized models perform better on repeated queries. However, the G-Click even performs worse than original ranking on non-repeated queries, which indicates the difficulty of inferring users' click behaviour on these queries. To tackle this issue, the SonetRank introduces social network and improves the results. The deep learning based model HTPS further enhances the performance by context-aware representation learning. Our model FNPS outperforms all baselines on both query sets, especially on non-repeated queries. This confirms that the group profile constructed by our model can accurately understand user query intent even if the query is a new one. 

\section{Conclusion}
In this work, we propose a group based personalized search model integrating search behaviour and friend network to refine the user profile. We combine the user's search behaviour and the friend network to form multiple friend circles and highlight similar users with neural networks. Under the interaction of relation-based and behaviour-based friend circles, the users who appear in both types of friend circles are enhanced to build the group profile. Finally, the search results can be personalized by combining the individual profile and the group profile with respect to the current query. Experimental results show a significant improvement of our model compared with existing personalized strategies.
\begin{acks}
Zhicheng Dou is the corresponding author. This work was supported by National Natural Science Foundation of China No. 61872370 and No. 61832017, Beijing Outstanding Young Scientist Program NO. BJJWZYJH012019100020098, Shandong Provincial Natural Science Foundation under Grant ZR2019ZD06, and the Outstanding Innovative Talents Cultivation Funded Programs 2020 of Renmin Univertity of China.
\end{acks}
\bibliographystyle{ACM-Reference-Format}
\balance
\bibliography{references}
\end{document}